\documentclass[pra,twocolumn,tightenlines,showpacs,nofootinbib]{revtex4}
\usepackage{bm,dcolumn,amsmath,graphicx}

\newcommand{\eref}[1]{Eq.~(\ref{#1})}
\newcommand{\tref}[1]{Table~\ref{#1}}

\begin{document}

\title{Binding energy and fission of the heavy charged massive particle -- nucleus bound state}
\author{V. V. Flambaum$^1$}
\author{G. McManus$^1$}
\author{S. G. Porsev$^{1,2}$}
\affiliation{$^1$ School of Physics, University of New South Wales,
Sydney, New South Wales 2052, Australia}
\affiliation{$^2$ Petersburg Nuclear Physics Institute, Gatchina,
Leningrad district, 188300, Russia}

\date{ \today }
\pacs{25.85.-w, 21.10.Dr, 95.35.+d}

\begin{abstract}
We consider a possibility of capture of a heavy charged massive particle $\chi^-$
by the nucleus leading to appearance of a bound state. A simple analytic
formula allowing to calculate binding energies of the $N\chi^-$ bound state for different nuclei
is derived. If the binding energy is sufficiently large the particle $\chi^-$ is
stable inside the nucleus. The probabilities of the nuclear fissions for such states are calculated.
It is shown that the bound states are more stable to a possible fission in comparison
to the bare nucleus. This makes an observation of this hypothetical charged massive
particle and the superheavy nuclei more probable.
\end{abstract}

\maketitle

%================================================================
\section{Introduction}
\label{sec_I}
%================================================================
The problem what the dark matter consists of is one of a hottest
topics of the modern physics. In particular, a possibility of
nonbaryonic dark matter is widely discussed (see, e.g., \cite{JunKamGri96,BerHooSil05}).
One of the possible models of the existence of nonbaryonic dark matter
was discussed by Pospelov and Ritz in their recent paper~\cite{PosRit08}.
They considered the direct
and indirect detection signatures of weakly interacting massive particles (WIMP)
$\chi^0$ with the mass $m_{\chi^0}$ with a heavier, but nearly degenerate, charged state $\chi^{\pm}$.
The WIMP-nucleus interaction may be dominated by inelastic
recombination process leading to the formation of $N\chi^-$ bound state.

Following this scenario and assuming that there is a heavy particle
with the mass $m_{\chi^-}$ and the charge equal to the electron charge $e$,
we derive a simple analytic formula allowing to calculate
binding energies of the $N\chi^-$ bound state for different nuclei.
This formula can be applied for any $Z$ and especially useful
for heavy and superheavy nuclei, whose properties are less studied so far.

In particular, we can consider the following
inelastic scattering of $\chi^0$ with heavy nucleus leading to the capture
of the particle $\chi^-$ by the nucleus and the appearance of the bound state
$N\chi^-$~\cite{PosRit08}:
\begin{equation}
\chi^0 + N \rightarrow (N\chi^-) + e^+ + \nu
\end{equation}
Knowing the binding energy $E_0$ we can predict how stable is the
bound state that occurred due to this $\beta^+$-type process.
The condition of stability of the particle $\chi^-$ inside the nucleus
is $|E_0| > m_{\chi^-} - m_{\chi^0}$.

The appearance of the bound state leads, in turn, to
that the probability of the tunneling through the fission barrier is changed.
From general considerations we can expect that the $N\chi^{-}$ bound state
will be more stable to a possible fission in comparison with the bare nucleus.
Our calculations show that for heavy nuclei the tunneling probability of the
bound state is reduced by a few orders of magnitude compared to the bare
nuclei. Since the bound states are more stable to a possible fission in comparison
to the bare nucleus, it opens new possibilities for the observations of such systems.

The paper is organized as follows.
In Sec.~\ref{sec_BE} we derive the analytic formula and calculate the binding
energies for different nuclei. In Sec.~\ref{sec_TP} we provide the expression
for the tunneling probability and describe the Coulomb interaction potential
for a deformed nucleus.  Section~\ref{sec_RD} is devoted to the discussion
of the obtained results and contains concluding remarks.
The units ($\hbar = c = 1$ and $e^2 \equiv \alpha \approx 1/137$)
are used throughout.

%================================================================
\section{Binding energy}
\label{sec_BE}
%================================================================
In this section we will find the binding energies of the $\chi^-$
particle to a nucleus. Pospelov and Ritz~\cite{PosRit08} numerically found the
binding energies of the state ($N\chi^{-}$) assuming a Gaussian
and steplike nuclear charge distribution for several elements and
solving the Schr\"{o}dinger equation with a given charge distribution
inside the nucleus. We derive analytical formulas which reproduce the numerical
results obtained in Ref.~\cite{PosRit08} to reasonable accuracy and allow us to calculate
the binding energies for heavy and superheavy nuclei.

Let us model a nucleus containing $Z$ protons as a uniformly charged sphere of radius $R$.
The electric potential is
\begin{equation}
V(r) = \left\lbrace
\begin{array}{l}
-\frac{Ze}{R} \left(\frac{3}{2} - \frac{1}{2}\frac{r^2}{R^2}\right), \,\,  0 < r \leq R, \\
-\frac{Ze}{r}, \,\, r \geq R ,
\end{array}
\right.
\label{Eq:V}
\end{equation}
where $r$ is the distance from the center of the nucleus.

Let us show that for heavy nuclei the wave function of the bound state $N\chi^{-}$
will be localized well within the nucleus. In other words,
the characteristic distances $r$ of the localization of the wave function
are smaller than the nuclear radius $R$ even for light nuclei.
For heavy nuclei $r \ll R$.

As follows from~\eref{Eq:V} (upper line) the second term of $V(r)$
is the potential of a three-dimensional harmonic oscillator.
The potential energy can be written as
\begin{equation}
U_{\rm osc}(r) = \frac{1}{2}\frac{Z \alpha \, r^2}{R^3} \equiv \frac{1}{2} M \omega^2 r^2 ,
\end{equation}
where we define
\begin{equation}
\omega = \sqrt{\frac{Z \alpha}{M R^3}}
\label{Eq:w}
\end{equation}
and $M \equiv m_{\chi^-} m_N/(m_{\chi^-} + m_N)$ is the reduced mass of
the particle $\chi^-$ and the nucleus. We assume that the mass
of the particle $m_{\chi^-}$ is comparable to or greater than the mass of the nucleus $m_N$.
If $m_{\chi^-} \gg m_N$ then $M\approx m_N$.

For the harmonic oscillator, as follows from the virial theorem,
the average potential and kinetic energies are equal
and for the total energy of the ground state we obtain
\begin{equation}
E_{\rm osc} = \frac{3}{2} \omega = M \omega^2 \langle r^2 \rangle.
\end{equation}
and, respectively,
\begin{equation}
\langle r^2 \rangle = \frac{3}{2 M \omega}.
\label{Eq:r2}
\end{equation}

Here and in the following it is sufficient for our purposes to
use the following approximations
\begin{equation}
R \approx r_0 A^{1/3} \,\, {\rm and} \,\, M \approx A\, m_p ,
\label{Eq:RM}
\end{equation}
where $r_0 \approx 1.2\,{\rm fm}$, $A$ is the nucleon number, and $m_p$ is the proton mass.

Using Eqs.~(\ref{Eq:w}), (\ref{Eq:r2}), and (\ref{Eq:RM}) and denoting
$r_{\rm av} \equiv \sqrt{\langle r^2 \rangle}$, we arrive at the following expression
for the ratio $r_{\rm av}/R$:
\begin{equation}
\frac{r_{\rm av}}{R} \approx \frac{2.7}{A^{1/3}\,Z^{1/4}}.
\end{equation}

In \tref{Tab:r_av} we present the results of calculation of $r_{\rm av}/R$ for
different elements. As seen from the table, for the heavy and the superheavy nuclei
$r/R \sim 0.1$. But even for the light nuclei we have $r_{\rm av} < R$.
% #####################################################################################################
\begin{table}
\caption{The ratio $r_{\rm av}/R$ for different nuclei.}
\label{Tab:r_av}
\begin{ruledtabular}
\begin{center}
\begin{tabular}{lrc}
$(^AN\chi^-)$               &  $Z$ & $r_{\rm av}/R$\\
\hline
$(^{11}\mathrm{B}\chi^-)$   &   5  & 0.81 \\
$(^{12}\mathrm{C}\chi^-)$   &   6  & 0.75 \\
$(^{14}\mathrm{N}\chi^-)$   &   7  & 0.69 \\
$(^{16}\mathrm{O}\chi^-)$   &   8  & 0.64 \\
$(^{40}\mathrm{Ar}\chi^-)$  &  18  & 0.38 \\
$(^{74}\mathrm{Ge}\chi^-)$  &  32  & 0.27 \\
$(^{132}\mathrm{Xe}\chi^-)$ &  54  & 0.20 \\
$(^{202}\mathrm{Hg}\chi^-)$ &  80  & 0.15 \\
$(^{232}\mathrm{Th}\chi^-)$ &  90  & 0.14 \\
$(^{257}\mathrm{Fm}\chi^-)$ & 100  & 0.13 \\
$(^{269}\mathrm{Ds}\chi^-)$ & 110  & 0.13
\end{tabular}
\end{center}
\end{ruledtabular}
\end{table}
% #####################################################################################################

Taking into account the estimate of the characteristic distance $r$,
we can neglect the wave function outside the nucleus and
approximate the potential energy using the electric potential inside the nucleus only.
This is the potential energy of a three-dimensional harmonic oscillator with the additional
constant term $-\frac{3}{2}\frac{Z\alpha}{R}$. It is given by
\begin{equation}
U(r) = -\frac{3}{2}\frac{Z\alpha}{R} + \frac{1}{2} M \omega^2 r^2 .
\end{equation}

The eigen-values of the Schr\"{o}dinger equation for the potential of a three-dimensional
harmonic oscillator are well-known. In our case there is the additional
constant term, so that the energies of this system are
\begin{equation}
E = -\frac{3}{2}\frac{Z\alpha}{R} + \omega \, (3/2 + n_x + n_y + n_z).
\label{Eq:E}
\end{equation}

Let us apply this model to the case of a nucleus with atomic weight $A$.
Using Eqs.~(\ref{Eq:w}) and (\ref{Eq:RM}) and writing the mass $M$ in units
of proton mass $m_p$ we obtain for the frequency:
\begin{equation}
\omega = \left( 7.73\,
\sqrt{\frac{Z}{A}\frac{m_p}{M}\frac{\mathrm{fm}^3}{r_0^3}} \right) \mathrm{MeV}.
\label{Eq:omega}
\end{equation}
The energy of the ground state is given by
\begin{equation}
E_0 = \left( -2.16\, \frac{Z}{A^{1/3}}\frac{\mathrm{fm}}{r_0}
 + 11.6\, \sqrt{\frac{Z}{A}\frac{m_p}{M}\frac{\mathrm{fm}^3}{r_0^3}} \right) \mathrm{MeV}.
\label{Eq:E00}
\end{equation}

As we have already mentioned the characteristic distance $r$ is much smaller than $R$ for heavy
nuclei. For this reason the contribution of the second term in~\eref{Eq:E00} is small for heavy
nuclei.

Taking into account \eref{Eq:RM} we arrive at the
following approximate formulas for the frequency $\omega$ and the ground state energy:
\begin{equation}
\omega \approx 5.9\, \frac{\sqrt{Z}}{A}\, \mathrm{MeV} ,
\label{Eq:w1}
\end{equation}
\begin{equation}
E_0 \approx \left( -1.8\, \frac{Z}{A^{1/3}} + 8.8\, \frac{\sqrt{Z}}{A} \right) \mathrm{MeV} .
\label{Eq:E0}
\end{equation}

%$r \sim (Z\alpha\,m_N)^{-1}$ and it can be readily shown that for heavy nuclei $r \ll R$.
%For instance, for $^{132}\!{\rm Xe}$ ($Z=54$) we obtain $r/R \sim 10^{-3}$.

In Table~\ref{Tab:BE} we list the frequencies and the binding energies calculated for several elements
using Eqs.~(\ref{Eq:w1}) and (\ref{Eq:E0}). As seen from the table there is a good agreement
between the energy values obtained in our simple approach and by Pospelov and Ritz~\cite{PosRit08}.
The heavier the atom the better agreement. A reason is that in our approach we assumed that
$r \ll R$ and, as a result, neglected the part of the Coulomb potential for $r > R$.
This approximation is good for the heavy elements while for the light
elements (like $^{11}\!B$) it can be used only for rough estimates.
% #####################################################################################################
\begin{table}
\caption{The frequencies $\omega$ (in MeV) and the binding energies (in MeV) of $\chi^-$
for different nuclei. The results are compared (where available) with those obtained
in Ref.~\cite{PosRit08}. If $|E_0|$ exceeds the mass difference $m_{\chi^-} - m_{\chi^0}$,
the particle $\chi^-$ is stable inside the nucleus.}
\label{Tab:BE}
\begin{ruledtabular}
\begin{tabular}{lcccc}
                \multicolumn{3}{c}{}              & \multicolumn{2}{c}{ $-E_0$ (MeV)} \\
   $(^A\!N\chi^-)$          & $Z$  &$\omega$ (MeV)& This work & Ref.~\cite{PosRit08} \smallskip \\
\hline \\
$(^{11}\mathrm{B}\chi^-)$    &   5  &     1.2      &   2.3     &  2.1 \smallskip \\
$(^{12}\mathrm{C}\chi^-)$    &   6  &     1.2      &   2.9     &  2.7 \\
$(^{14}\mathrm{N}\chi^-)$    &   7  &     1.1      &   3.6     &  3.2 \\
$(^{16}\mathrm{O}\chi^-)$    &   8  &     1.0      &   4.2     &  3.7 \\
$(^{40}\mathrm{Ar}\chi^-)$   &  18  &     0.63     &   8.5     &  8.0 \\
$(^{74}\mathrm{Ge}\chi^-)$   &  32  &     0.45     &  13.0     & 12.5 \\
$(^{132}\mathrm{Xe}\chi^-)$  &  54  &     0.33     &  18.6     & 18.4 \\
$(^{202}\mathrm{Hg}\chi^-)$  &  80  &     0.26     &  24.2     &   \\
$(^{232}\mathrm{Th}\chi^-)$  &  90  &     0.24     &  26.0     &   \\
$(^{257}\mathrm{Fm}\chi^-)$  & 100  &     0.23     &  28.0     &   \\
$(^{269}\mathrm{Ds}\chi^-)$  & 110  &     0.23     &  30.3     &   \\
%$(^{300}\mathrm{N_1}\chi^-)$ & 120  &     0.22     &           &   \\
%$(^{325}\mathrm{N_2}\chi^-)$ & 130  &     0.21     &           &   \\
%$(^{350}\mathrm{N_3}\chi^-)$ & 140  &     0.20     &           &   \\
%$(^{375}\mathrm{N_4}\chi^-)$ & 150  &     0.19     &           &   \\
%$(^{400}\mathrm{N_5}\chi^-)$ & 160  &     0.19     &           &
\end{tabular}
\end{ruledtabular}
\end{table}
% ######################################################################################################

% --------------------------------
\section{Tunneling probability}
\label{sec_TP}
% --------------------------------

In the work of Dzuba and Flambaum~\cite{DzuFla08} the effect of atomic
electrons on nuclear fission was considered. It was shown that atomic electrons
influence on the nuclear fission very insignificantly. The probability of the
fission of the nuclei with $Z \sim 100$ is changed only at the level of 0.2\%.

In this paper we consider effect of a heavy charged particle, which forms the bound state,
on the probability of the nuclear fission. Note that a real shape of the fission barrier
is complicated and accurate calculation requires the knowledge of the nuclear structure.
At the same time the qualitative features in the structure of the fission barrier can
be described by a simple parabolic barrier model~\cite{Seg77,BohMot74}.
Since our goal is to make an estimate of the nuclear fission probability, the
parabolic barrier model is sufficient for our purposes.

The probability of the tunneling through the barrier can be written as~\cite{BohMot74,DzuFla08}
% ------------------------------------------------------------------------------------
\begin{equation}
 P = \left[ 1+{\rm exp}\left( 2\pi \frac{|U_B - E|}{\omega_B} \right) \right]^{-1},
\label{Eq:P}
\end{equation}
% ------------------------------------------------------------------------------------
where $U_B$ is the maximum of the potential energy and
% the energy $E$ is assumed to be in vicinity of $U_B$.
the potential barrier width $\omega_B \sim 0.5 - 1 \, {\rm MeV}$~\cite{BohMot74}.

In a case of spontaneous fission we can estimate the difference $|U_B - E|$ for typical
energies $E$ as $|U_B - E| \sim 5$ Mev and therefore $2\pi |U_B - E|/\omega_B \gg 1$.
It is convenient to determine the probability of the spontaneous fission $P_0$ as
% -----------------------------------------------------------------
\begin{equation}
 P_0 = {\rm exp}\left(-2\pi \frac{|U_B - E|}{\omega_B} \right).
\label{Eq:P0}
\end{equation}
% -----------------------------------------------------------------
%This expression is valid if $E$ is not too close to $U_B$, so that
%$2\pi |U_B - E|/\hbar \omega \gg 1$.
Following Ref.~\cite{DzuFla08} we present the tunneling probability $P$ in the form
% ------------------------------------------------------------------------------
\begin{equation}
 P = P_0 \, {\rm exp} \left( -2\pi \frac{\delta E}{\omega_B} \right),
\label{P}
\end{equation}
% ------------------------------------------------------------------------------
where $\delta E \equiv E^{\rm C}_{max} - E^{\rm C}_{min}$ is the difference
between Coulomb energies of the particle $\chi^{-}$ in the points of maximum and minimum
of the nuclear energy $U$ as a function of the deformation parameter. To find
these Coulomb energies we have to specify the form of the nucleus. It will be discussed
in detail in the next section.
% --------------------------------
\subsection{Prolate ellipsoid}
\label{subsec_Pe}
% --------------------------------
Our subsequent calculations are based on the following experimental fact. The minimum
and maximum of the Coulomb energy correspond to spheroidal deformations
of the nucleus. Let us consider, for example, prolate spheroid with the
minor semiaxis $a$ and the major semiaxis $c$.
If this spheroid was obtained as a result of deformation of the sphere with
the radius $R$, then the condition of volume conservation reads as
% ------------------------------------------------------------------------------
\begin{equation}
 a^2 c = R^3 .
\end{equation}
% ------------------------------------------------------------------------------
If the eccentricity $\varepsilon$ is defined by
% ------------------------------------------------------------------------------
\begin{equation}
\varepsilon^2 = 1 - \frac{a^2}{c^2}
\label{e2}
\end{equation}
% ------------------------------------------------------------------------------
then for prolate ellipsoids
% ------------------------------------------------------------------------------
\begin{equation}
 a = R\,(1 - \varepsilon^2)^{1/6},
\end{equation}
\begin{equation}
 c = R\,(1 - \varepsilon^2)^{-1/3} .
\end{equation}
% ------------------------------------------------------------------------------
In Ref.~\cite{NilTsaSob69} (see also~\cite{LeaMol75}) it was introduced the parameter
deformation of the sphere $\eta$ connected with the minor and major semiaxes
of the prolate spheroid by a simple relation
% ------------------------------------------------------------------------------
\begin{equation}
 \frac{a}{c} = \frac{1 - \frac{2}{3} \eta}{1 + \frac{1}{3} \eta} .
\end{equation}
% ------------------------------------------------------------------------------
Using \eref{e2} we obtain for the eccentricity
% ------------------------------------------------------------------------------
\begin{equation}
 \varepsilon^2 = \frac{3\eta (6-\eta)}{(3 + \eta)^2} .
\end{equation}
% ------------------------------------------------------------------------------
% --------------------------------
\subsection{Coulomb potential}
\label{subsec_Cp}
% --------------------------------
The Coulomb potential of the spheroid can be written as~\cite{HasMye88}
% % ------------------------------------------------------------------------------
% \begin{equation}
%  U_{\rm C}(r,\theta) = -\frac{3}{2}\,\frac{Z e^2}{R}
% \left[ \frac{K}{\xi_0} - \frac{K-1}{\varepsilon^2}\,\frac{\xi^2}{\xi_0^2}
%     -\frac{1}{2 \varepsilon^2} \left( 1- \frac{K}{\xi_0^3}\right) \rho^2 \right] .
% \label{Phi1}
% \end{equation}
% % ------------------------------------------------------------------------------
% Here
% $$\xi \equiv \frac{r}{R}\, {\rm sin} \theta , \qquad
%  \rho \equiv \frac{r}{R}\, {\rm cos} \theta, \quad {\rm and} \quad
%  \xi_0 \equiv \frac{c}{R} ,$$

%
% If we take into account that $\xi_0 = (1-\varepsilon^2)^{-1/3}$, \eref{Phi1} can be rewritten as
% ------------------------------------------------------------------------------
\begin{eqnarray}
 U_{\rm C}(r,\theta) &=& -\frac{3}{2}\,\frac{Z \alpha}{R}  \left[ K (1-\varepsilon^2)^{1/3} + \right. \nonumber \\
&&\left.
\left\{ \frac{K-1}{\varepsilon^2}\,(\varepsilon^2 - 1)\, {\rm cos}^2 \theta \,+ \right. \right. \nonumber \\
&& \left. \left. \frac{1}{2} \left( \frac{K-1}{\varepsilon^2} - K \right) {\rm sin}^2 \theta \right\}
\frac{r^2}{R^2} \right] ,
\label{U_C}
\end{eqnarray}
% ------------------------------------------------------------------------------
 where the coefficient $K$ is determined as
% ------------------------------------------------------------------------------
\begin{equation}
 K \equiv \frac{1}{2\varepsilon}\, {\rm ln}\frac{1+\varepsilon}{1-\varepsilon}
\end{equation}
% ------------------------------------------------------------------------------
and $\theta$ is the angle between ${\bf r}$ and the axis $z$.

It is easy to show that if the eccentricity $\varepsilon \rightarrow 0$, i.e.
$a = c$, then
% ------------------------------------------------------------------------------
$$
K \rightarrow 1 \quad {\rm and} \quad \frac{K-1}{\varepsilon^2} \rightarrow \frac{1}{3} ,
$$
% ------------------------------------------------------------------------------
and the potential $U_{\rm C}$ goes over to the potential of the uniformly
charged sphere $U^{\rm sh}_{\rm C}$
% ------------------------------------------------------------------------------
\begin{eqnarray}
 U^{\rm sh}_{\rm C}(r,\theta) = -\frac{Z \alpha}{R}
 \left[ \frac{3}{2} - \frac{1}{2} \frac{r^2}{R^2} \right].
\label{U_sh}
\end{eqnarray}
% ------------------------------------------------------------------------------
% --------------------------------
\subsection{Deformation parameter}
\label{subsec_Dp}
% --------------------------------
According to experimental investigations (see, e.g.,~\cite{LeaMol75}),
in light actinides (like $^{236}$U) the deformation $\eta$ changes from
0.2 in the minimum to about 0.8 in the last maximum of the fission barrier.
In superheavy elements $\eta \simeq 0.4$ in the last maximum.

Using \eref{U_C} and supposing that in the minimum $\eta = 0.2$ and
in the maximum $\eta = 0.4$ we can find $\delta E = E^{\rm C}_{max} - E^{\rm C}_{min}$.
Taking into account that for heavy nuclei
the wave function of the bound state is localized at the distances
$r \ll R$, for an estimate we can neglect the term $\sim r^2/R^2$ in \eref{U_C}.
Then
% ------------------------------------------------------------------------------
\begin{eqnarray}
 U_{\rm C}(r,\theta) \approx  -\frac{3}{2}\,\frac{Z \alpha}{R} K (1-\varepsilon^2)^{1/3}
\label{U_C1}
\end{eqnarray}
% ------------------------------------------------------------------------------
The change of the parameter $\eta$ from 0.2 to 0.4 corresponds
to change of the eccentricity $\varepsilon$ from 0.58 to 0.76.
Then we obtain
% ------------------------------------------------------------------------------
\begin{eqnarray}
\frac{3}{2} K (1-\varepsilon^2)^{1/3} &=&
\frac{3\,(1-\varepsilon^2)^{1/3}}{4\varepsilon} {\rm ln}\frac{1+\varepsilon}{1-\varepsilon} \nonumber \\
&=& \left\lbrace
\begin{array}{l}
1.494,\, \varepsilon=0.58  \\
1.475,\, \varepsilon=0.76  \\
%1.355,\, \varepsilon=0.93  \\
\end{array}
\right.
\label{}
\end{eqnarray}
% ------------------------------------------------------------------------------
% For the atoms with very large $Z$:
% $E^{\rm C}_{\rm min}$ = $E^{\rm C}(\varepsilon = 0.2)$ and
% $E^{\rm C}_{\rm max}$ = $E^{\rm C}(\varepsilon = 0.4)$ we obtain
and, respectively,
% ------------------------------------------------------------------------------
\begin{eqnarray}
\left.
\begin{array}{l}
E^{\rm C}({\rm min}) \approx -1.494\, Z \alpha/R  \\
E^{\rm C}({\rm max}) \approx -1.475\, Z \alpha/R
\end{array}
\right\rbrace
\Rightarrow \delta E \approx 0.019 \, \frac{Z \alpha}{R} .
\label{Eq:E_C}
\end{eqnarray}
% ------------------------------------------------------------------------------
% ------------------------------
\section{Results and discussion}
\label{sec_RD}
% ------------------------------
Using the derived expressions for the minimal and maximal Coulomb energies
given by~\eref{Eq:E_C} we can find the tunneling probabilities for the heavy
and superheavy elements with $Z$ ranging from 80 to 160. For calculation we again
approximate the nuclear radius by $R \approx 1.2\, A^{1/3}\, {\rm fm}
\approx 0.006\, A^{1/3}\, {\rm MeV}^{-1}$.
For the superheavy nuclei with $Z \geq 120$ whose nucleon numbers $A$ are
not determined yet we put $A \approx 2.5\, Z$.
For an estimate we assume that $\omega_B \simeq 0.5 \, {\rm MeV}$~\cite{BohMot74}.

%Respectively, $R \approx 1.63\, Z^{1/3}\,{\rm fm}$ and
% Respectively, $R \approx 8.26\times 10^{-3}\, Z^{1/3}\,{\rm MeV}^{-1}$ and
% \begin{equation}
% \delta E \approx 0.02\, Z^{2/3}\, ({\rm MeV}) .
% \label{Eq:delE}
% \end{equation}
%
% Using~\eref{Eq:w1} we obtain for frequency in the same approximation
% \begin{equation}
% \omega \approx \frac{2.36}{\sqrt{Z}}\, (\mathrm{MeV}) ,
% \label{Eq:w2}
% \end{equation}

The results of calculation of $\delta E$ and $P/P_0$ are
listed in \tref{Tab:P}.
%Taking into account that $1 \, {\rm fm} = 1/(197.33 \, {\rm MeV})$,
%$e^2 = \alpha = 1/137$, and \eref{P} we find
% ####################################################################
\begin{table}[h]
\caption{The energy difference $\delta E$ (in MeV). The ratio of
the tunneling probability $P$ to the probability of
spontaneous tunneling $P_0$ for the elements with $Z$ ranging from 80 to 160
is calculated for $\omega_B = 0.5\, {\rm MeV}$.
The notation $y[x]$ means $y \times 10^x$.}
\label{Tab:P}
\begin{ruledtabular}
\begin{center}
\begin{tabular}{lccc}
%\hline \hline
  \multicolumn{1}{c}{$Z$}
& \multicolumn{1}{c}{$A$}
& \multicolumn{1}{c}{$\delta E$ (MeV)}
& \multicolumn{1}{c}{$P/P_0$} \\
\hline
\smallskip
 80  &  202  & 0.315 & 1.9[-2]   \\
\smallskip
 90  &  232  & 0.339 & 1.4[-2]   \\
\smallskip
100  &  257  & 0.364 & 1.0[-2]   \\
\smallskip
110  &  269  & 0.394 & 7.1[-3]   \\
\smallskip
120  &  300  & 0.414 & 5.5[-3]   \\
\smallskip
130  &  325  & 0.437 & 4.1[-3]   \\
\smallskip
140  &  350  & 0.459 & 3.1[-3]   \\
\smallskip
150  &  375  & 0.481 & 2.4[-3]   \\
\smallskip
160  &  400  & 0.501 & 1.8[-3]   \\
\end{tabular}
\end{center}
\end{ruledtabular}
%\footnotemark[1]{The nuclear radii are taken from Ref.~\cite{DzuFla08}}
\end{table}
% ####################################################################
Our calculation confirmed that the ($N\chi^{-}$) bound state
is more stable to a possible fission in comparison with the bare nucleus.
For the potential barrier width $\omega_B \simeq 0.5 \, {\rm MeV}$
the tunneling probability of the bound state are 2--3 orders of magnitude
smaller than that of the bare nucleus. The greater nuclear charge $Z$ the larger difference between
the probabilities $P$ and $P_0$. When $Z$ changes from 80 to 160 the
ratio $P/P_0$ changes from $1.9 \times 10^{-2}$ to $1.8 \times 10^{-3}$, i.e. it becomes
10 times smaller. It is not surprisingly because $P \sim \exp(-\delta_E)$ and
$\delta_E \sim Z/R $. Since $R \sim A^{1/3} \sim Z^{1/3}$,
$\delta_E$ increases as $Z^{2/3}$ with increasing $Z$.

%================================================================
% \section{Conclusion}
% \label{sec_C}
%================================================================
To conclude, we have considered the process of capture of the heavy charged massive particle $\chi^-$
by the nucleus leading to appearance of a bound state. We derived a simple analytic
formula allowing to calculate binding energies of the $N\chi^-$ bound state for different nuclei.
These energies can be calculated rather accurately for heavy and superheavy nuclei while
for light elements the derived formula can be used for the estimates of the
binding energies.

We have calculated the tunneling probabilities for a number of the $N\chi^-$ bound states
for the heavy and superheavy nuclei and showed that these states are more stable to a possible
fission in comparison to the bare nucleus. Their tunneling probabilities are 2--3 orders
of magnitude smaller than the tunneling probabilities of the bare nuclei. This result
is important because it opens new perspectives to observe such bound states and get
a new information about the hypothetical particle $\chi^-$ and the superheavy nuclei
which were not observed so far due to their instability.

%================================================================
% \section{Acknowledgments}
% \label{sec_Ac}
%================================================================
 This work was supported by
 the Australian Research Council. The work of S.G.P. was supported in part by
 the Russian Foundation for Basic Research under Grant No. 08-02-00460-a.

%###########################################################################################
%\bibliography{fission}

%###########################################################################################

\end{document}